\documentclass[twocolumn,showpacs,preprintnumbers,amsmath,amssymb]{revtex4}

\usepackage{graphicx}% Include figure files
\usepackage{dcolumn}% Align table columns on decimal point
\usepackage{bm}% bold math

\begin{document}

\preprint{APS/123-QED}

\title{Exotic Kondo-hole band resistivity and magnetoresistance of Ce$_{1-x}$La$_{x}$Os$_4$Sb$_{12}$ alloys}
\author{C. R. Rotundu}
 \altaffiliation[Now at ]{Department of Physics, University of Maryland, College Park, Maryland 20742-4111, USA}%Lines break automatically or can be forced with \\
\author{B. Andraka}%
 \email{andraka@phys.ufl.edu}
\affiliation{ Department of Physics, University of Florida\\
P.O. Box 118440, Gainesville, Florida  32611-8440, USA}%
\author{P. Schlottmann}%
\affiliation{ Department of Physics, Florida State University\\
Tallahassee, Florida  32306, USA}%

\date{\today}

\begin{abstract}
Electrical resistivity measurements of non-magnetic single-crystalline Ce$_{1-x}$La$_x$Os$_4$Sb$_{12}$ alloys, $x=0.02$ and 0.1, are reported for temperatures down to 20 mK and magnetic fields up to 18 T. At the lowest temperatures, the resistivity of Ce$_{0.98}$La$_{0.02}$Os$_4$Sb$_{12}$ has a Fermi-liquid-like temperature variation $\rho=\rho_0+A T^2$, but with negative $A$ in small fields. The resistivity has an unusually strong magnetic field dependence for a paramagnetic metal. The 20 mK resistivity increases by 75\% between $H=0$ and 4 T and then decreases by 65\% between 4 T and 18 T. Similarly, the $A$ coefficient increases with the field from -77 to 29$\,\mu\Omega$cmK$^{-2}$ between $H=0$ and 7 T and then decreases to 18$\,\mu\Omega$cmK$^{-2}$ for 18 T. This nontrivial temperature and field variation is attributed to the existence of a very narrow Kondo-hole band in the hybridization gap, which pins the Fermi energy. Due to disorder the Kondo-hole band has localized states close to the band edges. The resistivity for $x=0.1$ has a qualitatively similar behavior to that of $x=0.02$, but with a larger Kondo-hole band. 
\end{abstract}


\maketitle

\section{Introduction}

Lately, there has been considerable interest in materials belonging to the crystal structure of filled skutterudites\cite{Jeitschko}, with a general chemical formulae RT$_4$P$_{12}$, where R is a rare earth, T transition metal, and P pnictogen atom. This class of materials is unusually rich in exotic ground states and novel properties. These novel properties span heavy fermion superconductivity in PrOs$_4$Sb$_{12}$\cite{Bauer1}, a field-induced metal insulator-transition and heavy fermion behavior in PrFe$_4$P$_{12}$\cite{Aoki}, non-Fermi-liquid behavior in CeRu$_4$Sb$_{12}$\cite{Bauer2}, and possible Kondo insulator or Kondo semimetal behavior in CeOs$_4$Sb$_{12}$\cite{Bauer}. The ground state of the latter compound, a subject of our investigation, is still controversial.

The first experimental indication of a possible Kondo insulator ground state in CeOs$_4$Sb$_{12}$
was provided by the resistivity investigation of Bauer et al.\cite{Bauer} The resistivity has an activation-like temperature dependence between 10 and 40 K. The energy gap as measured by transport is small, about 10 K only. Much larger energy gaps of approximately 30 and 70 meV were more recently found in optical reflectivity (charge gaps)\cite{Matsunami} and inelastic neutron scattering measurements (spin gaps)\cite{Adroja}. It has been proposed that the 70 meV gap is the direct hybridization gap while the 30 meV gap is the indirect hybridization gap. However, no definite relationship between them has been established. In any case, the 10 K transport gap appears to be incompatible in size with those obtained from optical and neutron measurements. Furthermore, at temperatures smaller than 10 K, i.e., where a truly activation-type temperature dependence is anticipated for a semiconductor (assuming the 10 K gap), the rate of increase of the resitivity with a decrease of temperature is much smaller than expected. This low temperature resistivity is also strongly sample-dependent. The low temperature electronic specific heat coefficient is large and sample-dependent\cite{Bauer,Namiki,Rotundu}, valued between 90 and 180 mJ/K$^2$mol, and hardly consistent with the semiconducting gap at the Fermi level. 

These puzzling and somewhat contradictory properties posed the question whether extrinsic phases or crystal defects are responsible for the measured low temperature behaviors of CeOs$_4$Sb$_{12}$. The results of magnetic field investigations\cite{Namiki,Rotundu} argue against extrinsic phases responsible for the linear specific heat. 

In this work, we study how defects formed by the substitution of a small fraction (2-10\%) of Ce by La influence the electrical resistivity and magnetoresistance. It was previously demonstrated\cite{Rotundu}, that La-substitution is very effective in suppressing unconventional magnetic ordering taking place near 1 K. This low temperature magnetic behavior, although interesting in its own right, hinders the insight into the nature of the paramagnetic state of CeOs$_4$Sb$_{12}$. In addition to the corresponding anomaly at the transition temperature, it has been proposed\cite{Sugawara} that strong spin fluctuations persist at temperatures as high as 50 K and that these spin fluctuations would be responsible for deviations from semiconducting-like type temperature dependence of the electrical resistivity. 

\section{Experimental}

The samples used in this study were previously investigated\cite{Rotundu} by specific heat and electrical resistivity measurements down to 0.35 K, and magnetic susceptibility measurements to 1.8 K. 2\% of La was sufficient to suppress the 1 K anomaly to below 0.35 K. In this investigation we have extended the resistivity measurements down to 20 mK and have used a wide range of magnetic fields, between 0 and 18 T. In addition, the specific heat was measured down to 50 mK using a thermal relaxation method. However, the specific heat values are reliable down to 150 mK only, due to a strong nuclear quadrupolar contribution of Sb entering the measurement in a nontrivial manner\cite{Andraka}. The slow nuclear response of Sb at these low temperatures, characterized by time constants of the order of seconds and increasing with a decrease of temperature, resulted in large systematical errors at the lowest temperatures. Nevertheless, based on the specific heat measurement we can rule out any phase transition for Ce$_{0.98}$La$_{0.02}$Os$_4$Sb$_{12}$ to at least 100 mK. $C/T$ (specific heat over temperature) at 150 mK is $310 \pm 30$ mJ/K$^2$mol, which corresponds to a moderately heavy fermion system.

\begin{figure}
\includegraphics[width=3.3in]{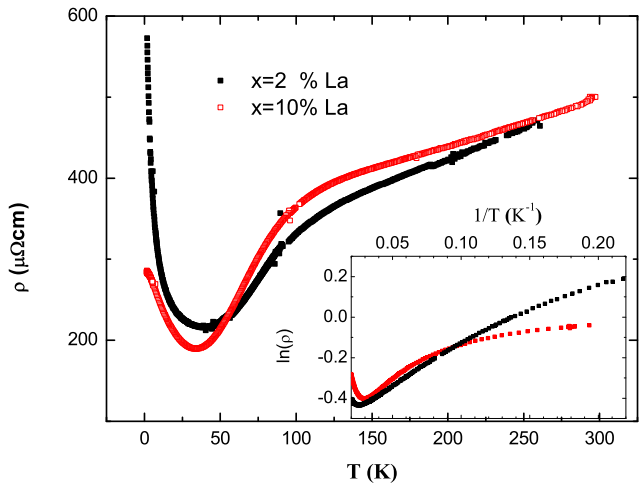} 
\caption{\label{fig:epsart} Electrical resistivity of Ce$_{1-x}$La$_{x}$Os$_4$Sb$_{12}$, $x=0.02$ and 0.1, between 1.5 and 300 K. The inset shows ln($\rho$) versus $1/T$. Color on-line.}
\end{figure}

\begin{figure}
\includegraphics[width=3.3in]{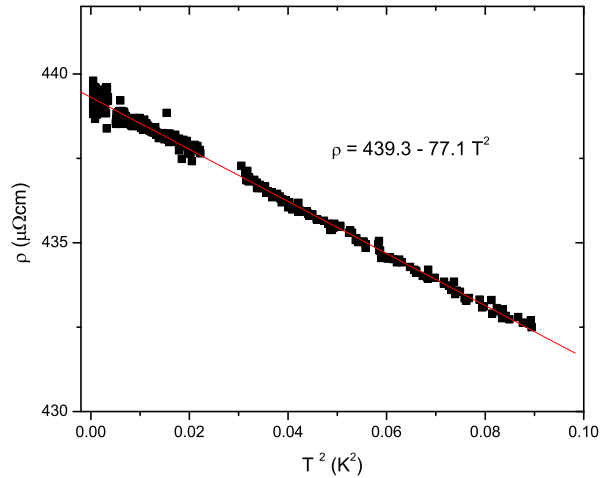} 
\caption{\label{fig:epsart} Electrical resistivity versus temperature square for Ce$_{0.98}$La$_{0.02}$Os$_4$Sb$_{12}$ between 20 and 300 mK in zero field. The solid line is a least square fit to $\rho=\rho_0+AT^2$.}
\end{figure}

\begin{figure}
\includegraphics[width=3.3in]{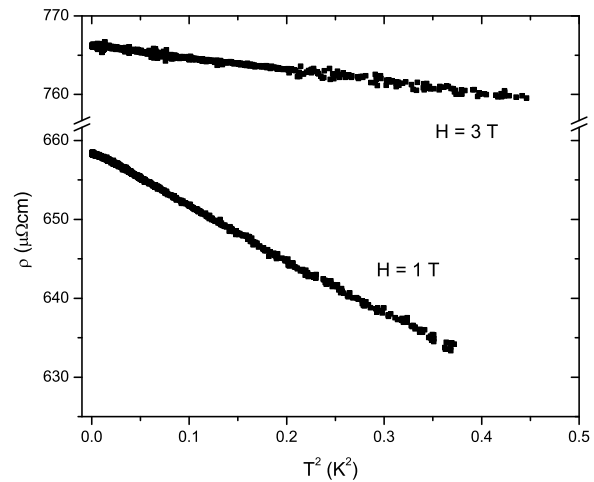} 
\caption{\label{fig:epsart} Electrical resistivity versus temperature square for Ce$_{0.98}$La$_{0.02}$Os$_4$Sb$_{12}$ in magnetic fields of 1 and 3 T.} \end{figure}

Similarly, no signatures of a phase transition were detected in the zero-field electrical resistivity of the investigated Ce$_{1-x}$La$_x$Os$_4$Sb$_{12}$ alloys, with $x=0.02$ and 0.1, down to 20 mK. Because of large uncertainties in geometrical factors, we relied on previous measurements in determining the absolute value of the resistivity. Both Bauer et al.\cite{Bauer} and Sugawara et al.\cite{Sugawara} reported the room temperature resistivity of CeOs$_4$Sb$_{12}$ to be approximately 500$\,\mu\Omega$cm. Therefore, we have assumed this value to be the room temperature resitivity of our $x=0.02$ and 0.1 crystals. This assumption has no bearing on the main conclusions of our study, derived from the temperature and magnetic field variations of the resistance. 

Figure 1 shows the electrical resistivity for $x=0.02$ and 0.1 between 1.5 K and room temperature. Both curves have a knee near 100 K, below which the resistivity sharply decreases before increasing again below 30 K. Interestingly, this reduction of electron scattering for $x=0.1$ is much more pronounced than in the previously studied pure compound or 2\% sample. The reduction of scattering below 100 K is consistent with a gradual formation of a hybridization gap or pseudogap. The charge fluctuations between the valence and conduction bands give rise to an enhanced density of states at the Fermi level and hence to an increased conduction. At lower $T$ the valence and conduction bands and the hybridization gap become better defined and the resistivity increases with decreasing $T$. The inset to Fig. 1 displays these resistivities between 5 and 50 K in a ln($\rho$) versus $1/T$ format. Clearly, the La-doping leads to a more metal-like resistivity.

Very interesting is the resistivity of the $x=0.02$ sample below 0.3 K (Fig. 2). The  temperature dependence can be approximated by the usual Fermi-liquid expression, $\rho=\rho_0+AT^2$, but with negative $A$ coefficient. We have observed a similar temperature dependence in other small fields (Fig. 3). However, as it can be inferred from Fig. 3, the absolute value of $A$ is strongly field-dependent. The absolute value of the coefficient $A$ decreases from about 77 to $15\,\mu\Omega$cmK$^{-2}$ between $H=0$ and 3 T. At the same time, the upper temperature limit for this temperature variation increases to approximately 0.6 - 0.7 K for $H=3$ T (from 0.3 K for $H=0$). 

\begin{figure}
\includegraphics[width=3.3in]{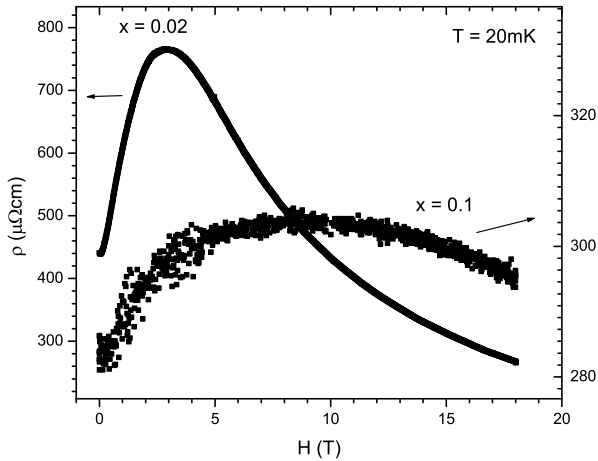}
\caption{\label{fig:epsart} Resistivity at 20 mK versus magnetic field for Ce$_{0.98}$La$_{0.02}$Os$_4$Sb$_{12}$ and Ce$_{0.9}$La$_{0.1}$Os$_4$Sb$_{12}$.} \end{figure}

Another unexpected result found for $x=0.02$ is a large increase of the low temperature resistivity with magnetic field (Figs. 3 and 4). This increase is inconsistent with the hybridization gap scenario, where we expect a decrease of the resistivity in large fields due to the closing of the hybridization gap. In pure CeOs$_4$Sb$_{12}$ there is indeed a strong reduction of the low temperature resistivity with magnetic field, to at least 14T\cite{Sugawara}. 
Above 2.9 T, up to at least 18 T, the resistivity at the lowest temperature (20 mK) decreases with the field strength (Fig. 4). It is worth emphasizing the large magnitude of this magnetoresistance. There is about an 80\% increase of $\rho$(20mK) between $H=0$ and 3 T and approximately 60\% decrease between $H=3$ and 18 T. 

The functional behavior of the resistivity changes near 4 T. For fields stronger than 4 T, the resistivity seems to follow the usual Fermi-liquid temperature dependence, with a positive temperature-square coefficient $A$ (Fig. 5). This temperature variation is only approximate. A closer inspection of the 7 T data suggests a slight upward curvature in $\rho$ versus $T^2$ at the lowest temperatures. Similarly small deviations from this Fermi-liquid temperature dependence can be detected for all other fields larger than 5 T. The $A$ coefficient, obtained from consistent least-square fits for all fields is shown in Fig. 6. This coefficient changes its sign from negative to positive near 4 T. 

The resistivity for more concentrated Ce$_{1-x}$La$_x$Os$_4$Sb$_{12}$ alloys appear to have qualitatively similar temperature-field variation to that for $x=0.02$. Fig. 4 shows the 20 mK magnetoresistance for $x=0.1$. The resistivity again initially increases with magnetic field, reaches a maximum value at approximately 9 T, which is a three times larger field than the one corresponding to the maximum for $x=0.02$, and decreases above 9 T. The relative changes of the resistivity in fields are much smaller (by a factor more than 10) than those for $x=0.02$. 
Also, the resistivity in zero field has the familiar temperature variation, $\rho=\rho_0+AT^2$, with $A \approx -0.5\,\mu\Omega$cmK$^{-2}$ (Fig. 7). Thus, the absolute value of $A$ is more than 100 times smaller than that for $x=0.02$. On the other hand, the range of temperatures for which this variation holds is much larger, up to approximately 5 - 6 K  (versus 0.6 K for $x=0.02$). 

\section{Discussion}

The temperature and magnetic field dependence of the resistivity for $x=0.02$ is very unusual. Particularly striking is the temperature variation in zero and small fields, i.e., $\rho=\rho_0+AT^2$, with a negative $A$. Such a temperature dependence has been previously reported for other Ce-alloys, such as Ce$_{0.97}$La$_{0.03}$Pd$_3$\cite{Lawrence}, and was attributed to Kondo holes. By definition a Kondo-hole is a non-$f$-electron impurity, such as La introduced for Ce in an otherwise periodic Ce-lattice. The Kondo-hole can behave similarly to a Kondo  impurity, this way strongly contributing to transport and thermodynamic properties at low temperatures\cite{Schlottmann,Schlottmann1}. The contribution of a Kondo-hole to the magnetoresistance should be similar to that of a Kondo-impurity, i.e., it should be negative. The magnetoresistance of the investigated Ce$_{1-x}$La$_x$Os$_4$Sb$_{12}$ alloys is negative only in sufficiently large fields, where $A$ is positive. Furthermore, there is an important difference in the way this negative $A$ coefficient develops upon La-doping in the investigated system and the aforementioned CePd$_3$. The low temperature resistivity of CePd$_3$ is dominated by the usual Coulomb scattering corresponding to a positive value of $A$. La substitution causes a decrease of $A$, which eventually turns negative. This is not what we observe in Ce$_{1-x}$La$_x$Os$_4$Sb$_{12}$. 

\begin{figure}
\includegraphics[width=3.3in]{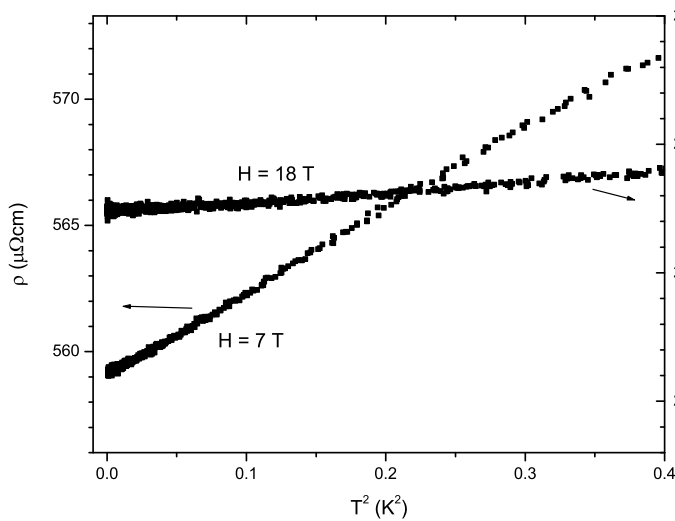} 
\caption{\label{fig:epsart} Electrical resistivity versus temperature square for Ce$_{0.98}$La$_{0.02}$Os$_4$Sb$_{12}$ in magnetic fields of 7 and 18 T.} \end{figure}

\begin{figure}
\includegraphics[width=3.3in]{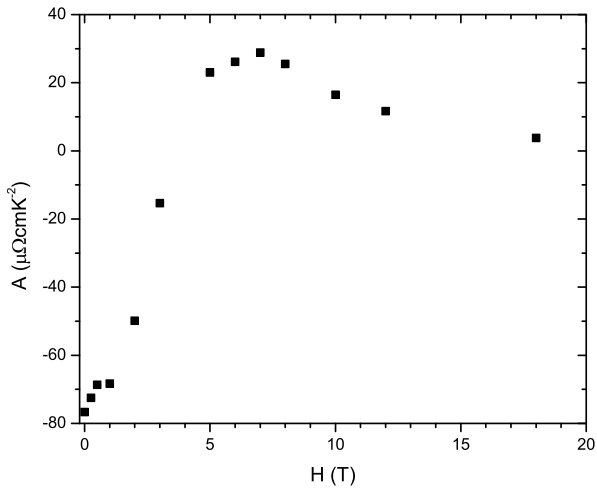}
\caption{\label{fig:epsart} Temperature-square coefficient ($A$) of the low temperature resistivity versus magnetic field for Ce$_{0.98}$La$_{0.02}$Os$_4$Sb$_{12}$.} 
\end{figure}

\begin{figure}
\includegraphics[width=3.3in]{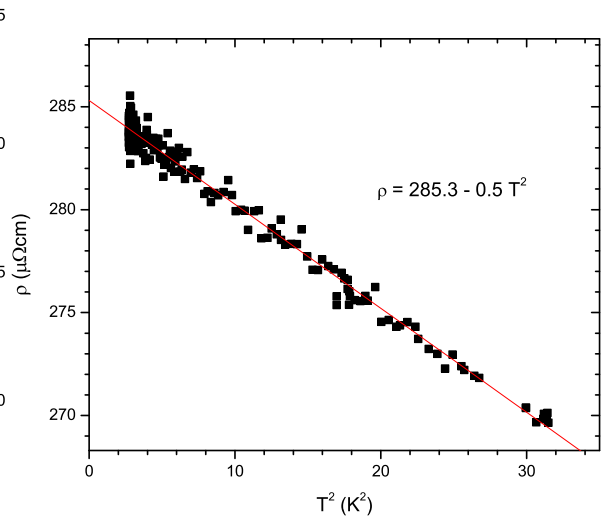} 
\caption{\label{fig:epsart} Electrical resistivity versus temperature square for Ce$_{0.9}$La$_{0.1}$Os$_4$Sb$_{12}$ between 1.5 and 6 K. The resistivity between 20 mK and 1.5 K (not shown due to large scattering) was consistent with the temperature dependence above 1.5 K expressed by the solid line, $\rho=\rho_0+AT^2$.}
\end{figure}

It is, therefore, important to consider the undoped material first. CeOs$_4$Sb$_{12}$ has clearly 
semiconducting-like character, which is not present in the resistivity of LaOs$_4$Sb$_{12}$. The semiconducting-like behavior is very strongly suppressed by minute amounts of La-impurities. Considering the close chemical similarity of La and Pr, the most natural explanation of such semiconducting-like behavior is then in terms of the hybridization gap. A hybridization gap would give rise to an exponential activation in the specific heat and not to a finite $\gamma$, i.e. a finite density of states at $E_F$, as observed experimentally. A finite $\gamma$ can, however, arise from localized states at the Fermi energy. Sample defects is the most likely mechanism leading to the electron localization. 

Several experimental observations, such as the sensitivity of the low temperature properties to sample preparation conditions or the forbidden indirect transition in optical measurements, strongly argue for a susceptibility of the material to sample defects. The fact that the variable range hopping mechanism is the dominant one for electronic transport under pressure\cite{Hedo} is also consistent with the scenario of localized states near $E_F$. Skutterudites are known to be susceptible to vacancies on the rare-earth sites. Pr-vacancies in PrOs$_4$Sb$_{12}$, of the order of a few percent, have been recently suggested\cite{Measson} to be responsible for the double superconducting transition in this material. A Ce-deficiency in CeOs$_4$Sb$_{12}$ of even a small fraction of a percent will dramatically affect the density of states near $E_F$ and the transport properties. 

Each missing Ce will give rise to a bound-state in the gap of the density of states (DOS) and for a fraction of a percent we would get an impurity band. Since the correlations at the vacancy site are weak, the impurity band should be closer to the valence band of the Kondo insulator than to the conduction band. In addition, these imperfections also affect the hybridization and a pseudogap (rather than a gap) can arise. For each Ce-vacancy we have 3 electrons missing. Hence, the Fermi level should lie in the valence band close to the gap-edge. The disorder due to randomly distributed Ce-vacancies also leads to a localization of the states in the pseudogap and in the valence band near $E_F$ and the gap-edge. The activation gap is then the energy difference between the Fermi level and the mobility edge, and hence much smaller than the hybridization gap. Variable range hopping is the most likely mechanism for electrical transport. This mechanism is then not intrinsic to the sample, but depends strongly on the sample preparation. A moderate Sommerfeld coefficient in the specific heat is expected from the localized correlated states at the Fermi level.

The substitution of Ce by La leads to a different type of defect, namely, the Kondo holes. Each La-ion contributes with one bound state in the gap (pseudogap). At a finite concentration (2\% or more La) the impurity bound states overlap and yield an impurity band (Kondo-hole band) roughly in the center of the gap, because the $f$-electrons are excluded from the La sites\cite{Schlottmann}. Assuming that the Ce vacancies are much less than 2\% (of La substitution) this impurity band is now the dominating one. In contrast to a Ce vacancy the La substitution is charge neutral. Hence, the Fermi level now lies in the impurity band and the system is expected to be metallic. The specific heat over $T$ is determined by the impurity band, i.e. it is concentration dependent and roughly corresponds to moderately heavy fermions.

Due to the disorder (Kondo holes and vacancies) not all the states in the impurity band are conducting. The impurity band has an upper and lower mobility edge and states closer to the band edges are localized. The states close to the Fermi level are conducting states. The system is then a Fermi liquid at low $T$. At intermediate $T$ carriers can be promoted from below the mobility edge and the conductivity is semiconductor-like. At even higher $T$ ($> 50$K) the hybridization gap is smeared and $\rho(T)$ is that of a bad metal.  

The temperature dependence of the resistivity is inversely proportional to the number of mobile carriers. At low $T$ the number of mobile carriers depends on the energy difference between the Fermi level and the mobility edge, i.e. the details of the density of states. In a Fermi liquid at low temperatures this energy difference should vary as $T^2$. Also the temperature should increase the number of mobile carriers, so that the conductivity (resistivity) increases (decreases) with $T^2$, in agreement with our observations. 

In zero-field and without vacancies the impurity band is half-filled. The magnetic field splits the Kondo-hole band via the Zeeman effect and shifts the up-spin and down-spin bands in opposite directions. This changes the number of carriers in the spin-subbands. While one subband becomes  electron-like (e.g. down-spin), the other is hole-like (up-spin). The carrier density in the impurity band is then effectively reduced by the magnetic field. Hence, the resistivity should increase with $H$, as seen in Figs. 3 and 4. The magnetic field also tends to reduce the spacial extension of the Kondo-hole bound states, i.e. favor localization, thus reducing the overlap between bound states. Hence, also the $A$ coefficient of the temperature dependence of the resistivity is expected to decrease with field.

%The carrier majority band becomes more and more metallic and for sufficiently strong fields the %effect of the chemical potential shift is reversed. This explanation is consistent with the %temperature and field variation of the resistivity of Ce$_{0.9}$La$_{0.1}$Os$_4$Sb$_{12}$, for %which we expect the Kondo-hole band a few times wider. 

\begin{acknowledgments}
Work supported by the Department of Energy, grant No. DE-FG02-99ER45748, and the National High Magnetic Fields Laboratory. P.S. is supported by the Department of Energy, grant No. DE-FG02-98ER45707.
\end{acknowledgments}

\bibliography{rho1}% Produces the bibliography via BibTeX.

\end{document}